\documentclass[12pt]{article}
\usepackage{natbib}
\usepackage{apalike}
\usepackage{epsfig}
\usepackage{setstack}
\usepackage{lscape}
\usepackage{wrapfig}
\usepackage{times}
\usepackage{geometry}
\usepackage{wasysym}
\usepackage[utf8]{inputenc}
\usepackage{enumitem,amssymb}
\usepackage{ragged2e}
\newlist{thematic}{itemize}{8}
\setlist[thematic]{label=$\square$}
\usepackage{pifont}

\begin{document}
\raggedright
\huge
Astro2020 APC White Papers \linebreak

A Realistic Roadmap to Formation Flying Space Interferometry \linebreak
\normalsize

\noindent \textbf{Type of Activity:} \hspace*{60pt} $\square$ Ground Based Project \hspace*{10pt} $\XBox$ Space Based Project \hspace*{20pt}\linebreak
$\square$ Infrastructure Activity \hspace*{31pt} $\XBox$ Technological Development Activity \linebreak
  $\square$  State of the Profession Consideration \hspace*{1pt} $\square$ Other {\underline{~~~~~~~~~~~~~~~~~~~~~~~~~~~~~~~~~~~~~~~~~~~~}} \hspace*{40pt} \linebreak
 \linebreak
  
\textbf{Principal Author:}

Name: John Monnier	
 \linebreak						
Institution:  University of Michigan
 \linebreak
Email: monnier@umich.edu    
 \linebreak
Phone:  734-763-5822
 \linebreak
 
\textbf{Endorsers:} (names, institutions, emails)
  \linebreak
Alicia	Aarnio,	University of North Carolina Greensboro,		anaarnio@uncg.edu		\\
Olivier	Absil,	University of Liège,		olivier.absil@uliege.be		\\
Narsireddy	Anugu,	University of Exeter,		N.anugu@exeter.ac.uk		\\
Ellyn	Baines,	Naval Research Lab,		ellyn.baines@nrl.navy.mil		\\
Amelia	Bayo,	Universidad de Valparaiso,		amelia.bayo@uv.cl		\\
Jean-Philippe	Berger,	Univ. Grenoble Alpes, IPAG,		jean-philippe.berger@univ-grenoble-alpes.fr		\\
L. Ilsedore	Cleeves,	University of Virginia,		lic3f@virginia.edu		\\
Daniel	Dale,	U. Wyoming,		ddale@uwyo.edu		\\
William	Danchi,	NASA Goddard Space Flight Center,		william.c.danchi@nasa.gov		\\
W.J.	de Wit,	ESO,		wdewit@eso.org		\\
Denis	Defrère,	University of Liège,		ddefrere@uliege.be		\\
Shawn	Domagal-Goldman,	NASA-GSFC,		shawn.goldman@nasa.gov	\\
Martin Elvis, CfA Harvard \& Smithsonian, melvis@cfa.harvard.edu \\
Dirk	Froebrich,	University of Kent,		df@star.kent.ac.uk		\\
Mario	Gai,	Istituto Nazionale di AstroFisica,,		mario.gai@inaf.it		\\
Poshak	Gandhi,	School of Physics \& Astronomy,		poshak.gandhi@soton.ac.uk		\\
Paulo	Garcia,	Universidade do Porto, Portugal,		pgarcia@fe.up.pt		\\
Tyler	Gardner,	University of Michigan,		tgardne@umich.edu		\\
Douglas	Gies,	Georgia State University,		gies@chara.gsu.edu		\\
Jean-François	Gonzalez,	Univ Lyon,		Jean-Francois.Gonzalez@ens-lyon.fr		\\
Brian	Gunter,	Georgia Institute of Technology,		brian.gunter@aerospace.gatech.edu		\\
Sebastian	Hoenig,	University of Southampton, UK,		s.hoenig@soton.ac.uk		\\
Michael	Ireland,	Australian National University,		michael.ireland@anu.edu.au		\\
Anders 	Jorgensen,	New Mexico Institute of Mining \& Tech,		Anders.M.Jorgensen@nmt.edu		\\
Makoto	Kishimoto,	Kyoto Sangyo University, Japan,		mak@cc.kyoto-su.ac.jp		\\
Lucia	Klarmann,	Max-Planck-Institut für Astronomie,		klarmann@mpia.de		\\
Brian	Kloppenborg,	Georgia Tech Research Institute,		brian.kloppenborg@gtri.gatech.edu		\\
Jacques	Kluska,	KU Leuven,		jacques.kluska@kuleuven.be		\\
J Scott	Knight,	Ball Aerospace \& Technologies, 		jsknight@ball.com		\\
Quentin	Kral,	Paris Observatory/Lesia,		quentin.kral@obspm.fr		\\
Stefan	Kraus,	University of Exeter,		s.kraus@exeter.ac.uk		\\
Lucas	Labadie,	University of Cologne, Germany,		labadie@ph1.uni-koeln.de		\\
Peter	Lawson,	Caltech, Jet Propulsion Laboratory,		peter.r.lawson@jpl.nasa.gov		\\
Jean-Baptiste	Le Bouquin,	University of Michigan,		lebouquj@umich.edu		\\
David	Leisawitz,	NASA GSFC,		david.t.leisawitz@nasa.gov		\\
E. Glenn	Lightsey,	Georgia Institute of Technology,		glenn.lightsey@gatech.edu		\\
Hendrik	Linz,	MPIA Heidelberg,		linz@mpia.de		\\
Sarah	Lipscy,	Ball Aerospace		slipscy@ball.com		\\
Meredith	MacGregor,	Carnegie DTM,		mmacgregor@carnegiescience.edu		\\
Hiroshi	Matsuo,	National Astronomical Observatory of Japan,		h.matsuo@nao.ac.jp		\\
Bertrand	Mennesson,	JPL,		bertrand.mennesson@jpl.nasa.gov		\\
Michael	Meyer,	University of Michigan,		mrmeyer@umich.edu		\\
Ernest A.	Michael,	University of Chile,		ernest.michael@raig.uchile.cl		\\
Florentin	Millour,	Université Côte d'Azur,		florentin.millour@oca.eu		\\
David	Mozurkewich,	Seabrook Engineering,		dave@mozurkewich.com		\\
Ryan	Norris,	Georgia State University,		norris@astro.gsu.edu		\\
Marc	Ollivier,	Institut d'Astrophysique Spatiale Orsay,		marc.ollivier@ias.u-psud.fr		\\
Chris	Packham,	UTSA,		chris.packham@utsa.edu		\\
Romain	Petrov,	Côte d'Azur University,		romain.petrov@unice.fr		\\
Benjamin	Pope,	New York University,		benjamin.pope@nyu.edu		\\
Laurent	Pueyo,	STScI,		pueyo@stsci.edu \\
Sascha	Quanz,	ETH Zurich,		sascha.quanz@phys.ethz.ch		\\
Sam	Ragland,	W.M. Keck Observatory,		sragland@keck.hawaii.edu		\\
Gioia	Rau,	NASA/GSFC,		gioia.rau@nasa.gov		\\
Zsolt	Regaly,	Konkoly Observatory,		regaly@konkoly.hu		\\
Alberto	Riva,	INAF - Osservatorio Astrofisico di Torino,		alberto.riva@inaf.it		\\
Rachael	Roettenbacher,	Yale University,		rachael.roettenbacher@yale.edu		\\
Giorgio	Savini,	University College London,		g.savini@ucl.ac.uk		\\
Benjamin	Setterholm,	University of Michigan,		bensett@umich.edu		\\
Marta	Sewilo,	NASA/GSFC, University of Maryland,		marta.m.sewilo@nasa.gov		\\
Michael	Smith,	University of Kent,		m.d.smith@kent.ac.uk		\\
Locke	Spencer,	University of Lethbridge,		Locke.Spencer@uLeth.ca		\\
Theo	ten Brummelaar,	CHARA - Georgia State University,		theo@chara-array.org		\\
Neal	Turner,	Jet Propulsion Laboratory, Caltech,		neal.turner@jpl.nasa.gov		\\
Gerard	van Belle,	Lowell Observatory,		gerard@lowell.edu		\\
Gerd	Weigelt,	MPI for Radio Astronomy,		weigelt@mpifr.de		\\
Markus	Wittkowski,	ESO,		mwittkow@eso.org		\\

\pagebreak
\justifying

\noindent \textbf{Abstract:}

The ultimate astronomical observatory would be a formation flying interferometer in space, immune to atmospheric turbulence and absorption, free from atmospheric and telescope thermal emission, and reconfigurable to adjust baselines according to the required angular resolution.  Imagine the near/mid-infrared  sensitivity of the JWST and the far-IR sensitivity of Herschel but with ALMA-level angular resolution, or imagine having the precision control to null host star light across 250m baselines and to detect molecules from the atmospheres of nearby exo-Earths.  With no practical engineering limit to the formation's size or number of telescopes in the array, formation flying interferometry will revolutionize astronomy and this White Paper makes the case that it is now time to accelerate investments in this technological area.  Here we provide a brief overview of the required technologies needed to allow light to be collected and interfered using separate spacecrafts.  We emphasize the emerging role of inexpensive smallSat\footnote{Here, smallSats are satellites with mass $<$180kg, \\ including nanosats, cubesats, microsats, minisats, etc.} projects and the excitement for the LISA Gravitational Wave Interferometer to push development of the required engineering building-blocks. {\bf We urge the Astro2020 Decadal Survey Committee to highlight the need for a small-scale formation flying space interferometer project to demonstrate end-to-end competency with a timeline for first stellar fringes by the end of the decade.}

\vspace{0.1in}

\begin{wrapfigure}[27]{r}{0.5\textwidth}
\vspace{-.5in}
\begin{center}
   \begin{tabular}{c} %
   \includegraphics[width=3.18in]{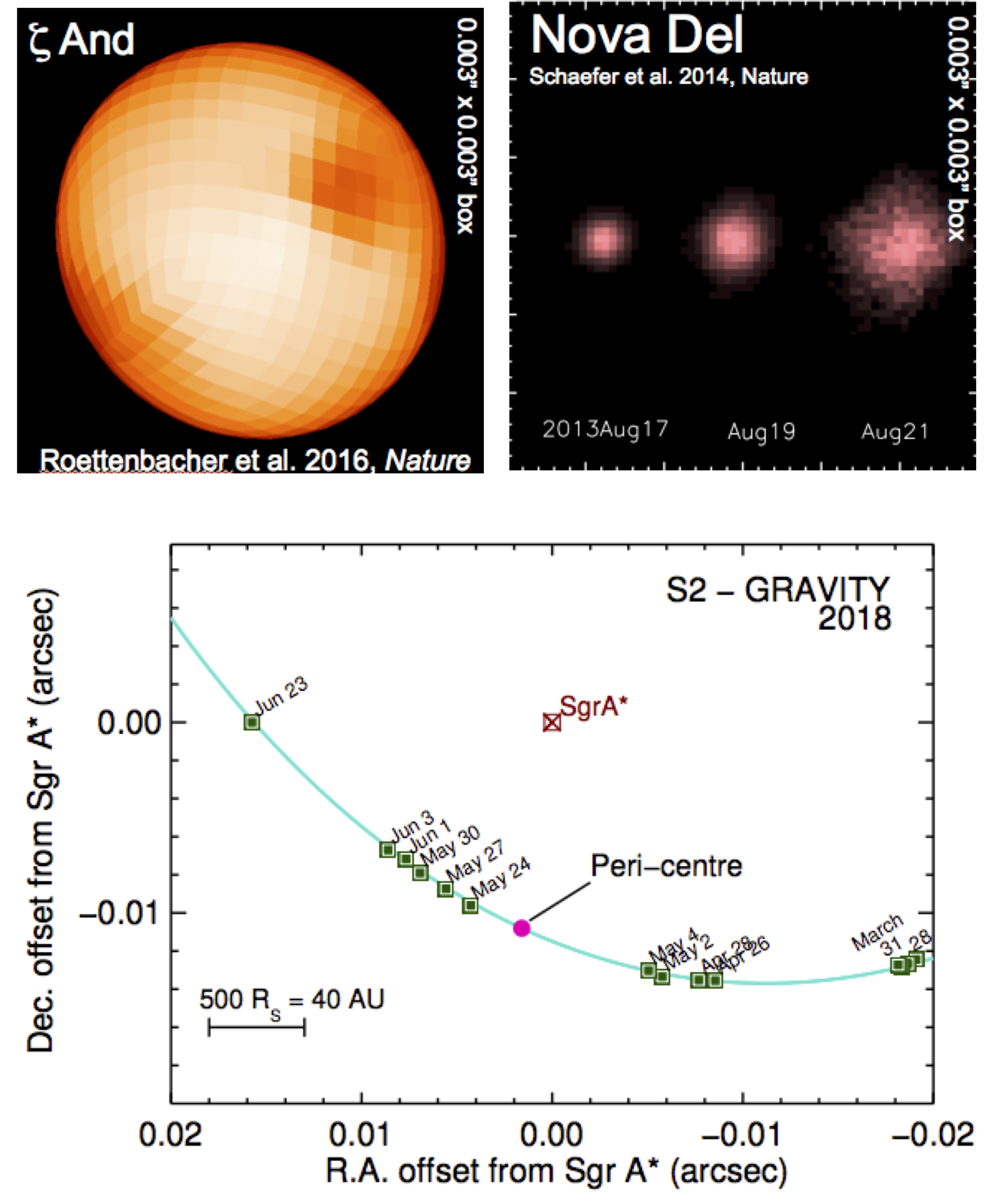}
   \end{tabular}
   \end{center}
   \vspace{-.3in}
   \caption[example] 
   { \label{bestresults} 
\footnotesize Ground-based infrared interferometry has come a long way in the past 20 years. Here we see some recent imaging results from \citet{roettenbacher2016}, \citet{schaefer2014}, and \cite{abuter2018}, spanning stellar imaging, novae explosions, and stellar orbits around Sgr~A* that probe General Relativity. Note that the top two image boxes are only 3 {\em milli-arcseconds} across while the stellar orbit shown has errors  $<$100 {\em micro-arcseconds}.}
   \end{wrapfigure} 
   
\noindent {\bf Caveats:} The landscape of formation flying space missions is rapidly developing.   In addition, some important relevant past missions are not well known and often were only sparsely described with few published results.  The authors regret the inevitable incompleteness in trying to summarize an emerging field -- please contact monnier@umich.edu with corrections and additions.

\vspace{0.1in}

\noindent {\bf Key Science Goals and Objectives:}

Long-baseline interferometry in the infrared and visible from the ground\ is now routine, obtaining images of stars and their environments with milli-arcsecond angular resolution, for instance rapidly-rotating stars \citep{monnier2007}, magnetically-active stars \citep{roettenbacher2016}, stars zipping around our galaxy's supermassive black hole \citep{abuter2018}, galactic Novae exploding in realtime \citep{schaefer2014},  a quasar's broad-line region \citep{sturm2019}, and even exoplanet spectra directly \citep{lacour2019}. See Figure~\ref{bestresults} for a gallery of recent interferometric imagery.

\begin{wrapfigure}[21]{r}{3in}
\vspace{-.2in}
\begin{center}
   \begin{tabular}{c} %
   \includegraphics[width=3in]{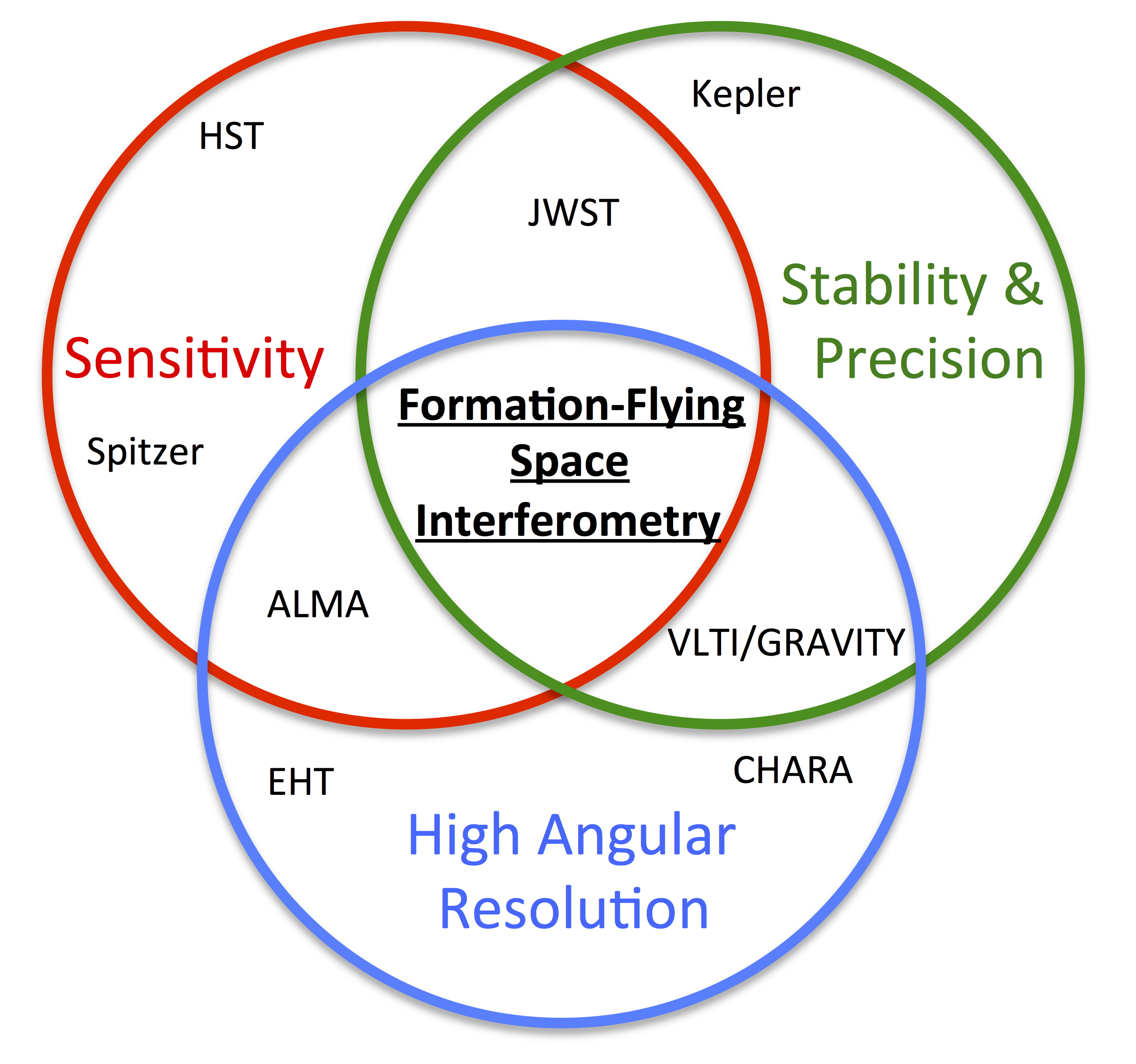}
   \end{tabular}
   \end{center}
   \vspace{-.3in}
   \caption[example] 
   { \label{venn} 
\footnotesize Space interferometry represents the ultimate astronomical observatory, capable of combining the power of sensitivity, stability and high angular resolution.  Some of our most exciting astronomical goals require all three, namely mid-infrared exoplanet characterization using nulling or high-resolution image of high-redshift galaxies. }
   \end{wrapfigure} 
   
Despite these remarkable achievements, the sensitivity of this technique\footnote{We are specifically considering ``direct detection'' interferometry where photons have to be directly interfered using coherent beam transport using mirrors or single-mode fibers.  Radio and mm-wave interferometry can use heterodyne detection and is far less limited by the atmosphere.}  is harshly limited by atmospheric turbulence -- which restricts coherent integration times to $<<1$~second  -- and by the high price of large aperture telescopes -- generally limiting most visible/IR interferometer facilities to $<1.5$m telescopes presently. For example, the CHARA interferometer has a limiting magnitude K$\sim$8 with 1.0m telescopes in the near-IR while the VLTI can reach K$\sim$10 with the 8m UT telescopes\footnote{VLTI/GRAVITY can observe K$\sim$17 mag objects, such as flares around Sgr~A*, as long as there is a bright phase reference source within a few arcseconds of the fainter target.}.  
ALMA has shown us the power of combining the high angular resolution imaging of interferometry with a revolutionary boost in sensitivity.  Ground-based facilities have matured to the point that planning can begin for the next generation of capabilities.

Figure\,\ref{venn} shows a simple reminder of why the allure of space interferometry is so great.  The advantages of space over ground are fully realized:  Sensitivity, Stability, and High angular resolution.  
The gains will be particularly dramatic in the mid- and far-infrared where existing techniques have left a huge gap in angular resolution and sensitivity.

\begin{wrapfigure}[18]{l}{4in}
\vspace{-.4in}
\begin{center}
   \begin{tabular}{c} %
   \includegraphics[width=4in]{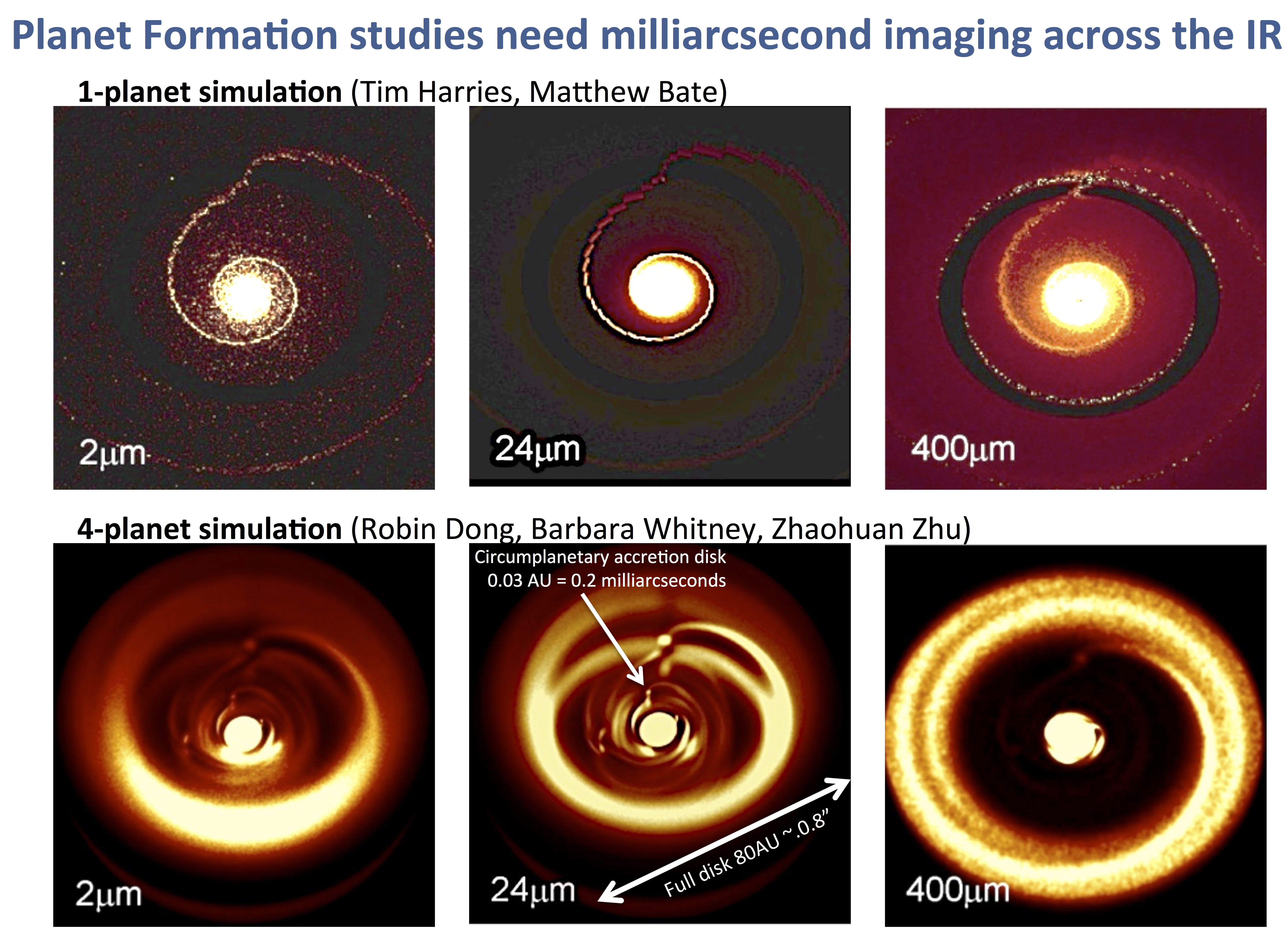}
   \end{tabular}
   \end{center}
   \vspace{-.3in}
   \caption[example] 
   { \label{sims} 
\footnotesize Multi-wavelength, milli-arcsecond resolution imaging will be needed to untangle the complex physical processes at play during planet formation, as this suite of simulations shows.  Even more imaging complexity will likely be seen for extragalactic sources. }
   \end{wrapfigure}

Planet-forming disks, for example, show a high level of complexity that we hope to understand with high resolution imaging across the IR.  Figure~\ref{sims} shows examples from a suite of simulations run as part of the Planet Formation Imager project studies \citep{monnier_pfi2018}.  We anticipate that high-redshift galaxies will also show a high level of structural complexity and a formation flying space interferometer would be a very powerful follow-up to expected exciting results from JWST.

Imaging complex sources such as galaxies and protoplanetary disks will require at least a dozen (or more) telescopes in the interferometric array to synthesize complete (u,v)-coverage -- recall that ALMA routinely uses $\sim$50 antennas to create the breathtaking images of protoplanetary disks \citep[e.g., D-SHARP;][]{andrews2018}.  A simpler space interferometer (e.g., LIFE, Large Interferometer for Exoplanets; https://www.life-space-mission.com/)  with just 4x2.8m space telescopes spread over 500m could use nulling mid-IR interferometry for exoplanet detection and spectral characterization. Armed with decent knowledge of exoplanet demographics from Kepler, \citet{kammerer2018} found that such an interferometer would detect hundreds of small planets.   The detection statistics are very favorable and would highly complement the visible light exoplanet detections from a future HabEx or LUVOIR coronagraphy mission.  

{\em Please see additional relevant White Papers submitted to the ASTRO2020, including ``Imaging the Key Stages of Planet Formation (lead: Monnier)'', ``The Future of Exoplanet Direct Detection (lead: Monnier)'', and ``A Long-Term Vision for Space-Based Interferometry'' (Lead: Rinehart). }

\vspace{.1in}
   
\noindent {\bf Technical Overview:}

One can trace formation flying space interferometry proposals back to the 1970s and 1980s, including early work on nulling \citep{bracewell1979}, a 3-telescope mission concept \citep[TRIO;][]{labeyrie1984,labeyrie1985} and some interesting orbital mechanics considerations for low-earth orbits  \citep{stachnik1985}.  Ultimately the first optical interferometer in space was flown as part of a single telescope -- the Hubble Space Telescope actually employs  interferometers within the Fine Guidance Sensors \citep{eaton1993}.

\begin{wraptable}[20]{r}{4.2in}
\vspace{-.4in}
\footnotesize
\begin{center}
\caption{Proposed space interferometer concepts (NASA/ESA) }
\begin{tabular}{|l|l|l|}
\hline
Name & Wavelength & Reference \\
\hline
SIM & visible & \citet{shao1998} \\
OSIRIS & visible & \citet{bagrov1999} \\
MAXIM & X-ray & \citet{cash2004} \\
MAXIM Pathfinder & X-ray & \citet{gendreau2004} \\
PEGASE & vis/NIR & \citet{leduigou2006} \\
SPECS & far-IR & \citet{harwit2006} \\	
ESPRIT & far-IR & \citet{wild2008} \\		
SPIRIT	& far-IR & \citet{leisawitz2009}\\ 	
FIRI & far-IR & \citet{helmich2009} \\			
DARWIN & mid-IR & \citet{cockell2009} \\
SIM-LITE & visible & \citet{shao2010} \\
FKSI & mid-IR &  \citet{danchi2010} \\
Stellar Imager & UV/visible & \citet{carpenter2010}\\
TPF-I & mid-IR & \citet{martin2011}\\
SHARP-IR & far-IR & \citet{rinehart2016} \\
DARE & radio & \citet{plice2017} \\
LISA & n/a & \citet{lisa2017} \\
SunRISE & radio & \citet{lazio2018} \\
LIFE & mid-IR & \citet{quanz2018} \\
IRASSI & far-IR & \citet{linz2019} \\
\hline
\end{tabular}
\label{missions}
\end{center}
\vspace{-.2in}
\end{wraptable}

The first major push for a dedicated space interferometer  was for {\bf astrometry}, as atmospheric turbulence fundamentally limits the prospects for parallax and proper motions measurements on the full sky.  
Following encouragement from the 1990 Decadal survey, the Space Interferometry Mission \citep[SIM,][]{shao1998} was developed by NASA, with a goal for 4$\mu$-arcsecond astrometric precision using a 10m baseline on a structurally-connected interferometer\footnote{While there may be a short-term case for structurally-connected space interferometers, we are focused here in developing rationale for formation flying interferometry which is more promising the long-term.}.  Due to technical delays, rising costs, and emerging competing techniques, SIM was cancelled as part of the Astro2010 Decadal Survey.
{\em High-precision astrometry using interferometers is technically difficult and has been largely abandoned in favor of more traditional architectures, relaxing stability constraints from nm-level to $\mu$m level.}  

Moving on, the next application explored seriously for space interferometry was {\bf nulling for exoplanet detection}.  Interferometric nulling \citep{bracewell1979} is the technique whereby the interferometer is tuned so that the host star sits at a destructive interference ``null'' while nearby exoplanets would generally not be nulled and thus be detected.  Because light from the star is suppressed, there is no associated photon noise.  Nulling has limited application on the ground due to the strong piston fluctuations induced by atmospheric turbulence that frustrate maintenance of the destructive null.  In the 2000s, the NASA Terrestrial Planet Finder Interferometer \cite[TPFI,][]{lawson2007} and ESA DARWIN \citep{cockell2009} projects studied the requirements for a mid-IR formation flying nulling space interferometer. The core idea is that an exo-Earth has the best contrast with the host star in the mid-IR (1:10$^6$, compared to 1:10$^{10}$ in visible) and nulling will further suppress the host star flux.    In mid/late 2000s, these technology projects failed to mature to funded flight missions -- see \citet{lawson2009} and \citet{lawson2017} for a timeline of events and an overview of technical achievements of the TPFI program.

Additional  mission concepts have been developed for nearly all wavelengths ranges and also using unconventional methods such as intensity interferometry \citep{matsuo2018}.  A (nearly) exhaustive list of proposed major space interferometry mission concepts appears in Table~\ref{missions}.

\begin{figure*}[!b]
\begin{center}
   \begin{tabular}{c} %
   \includegraphics[height=3.5in]{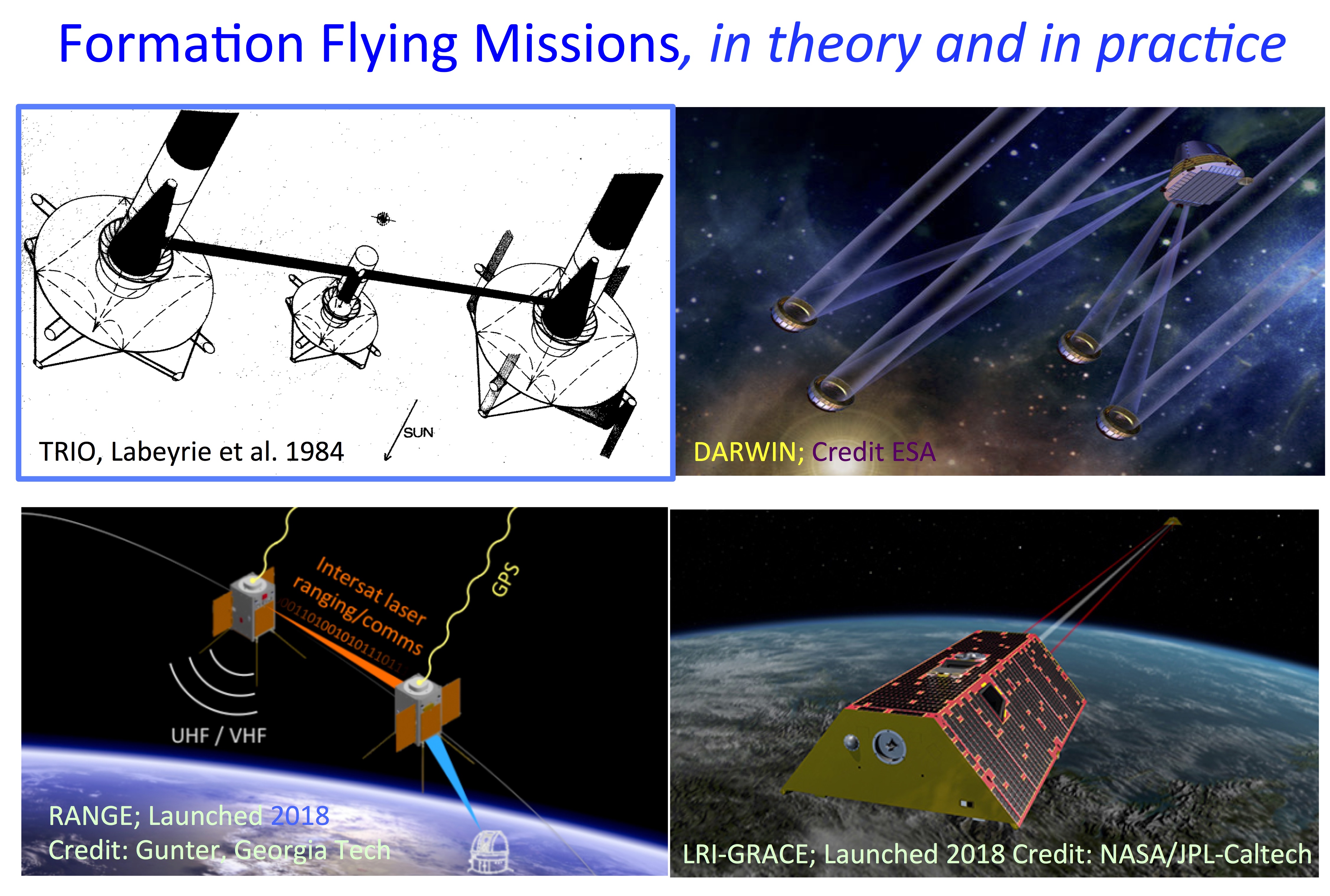}
   \end{tabular}
   \end{center}
   \vspace{-.3in}
   \caption[example] 
   { \label{gallery} 
\footnotesize Astronomers have dreamed big for decades but are stalled in moving forward on ambitious space interferometers (top panels). Major breakthroughs in formation flying  have been achieved by other communities (bottom panels), reinvigorating work towards true formation flying space interferometer with first stellar fringes by 2030.  }
   \end{figure*}

Since cancellation of SIM, TPFI, and DARWIN, there has been only limited organized activity within NASA or ESA developing space interferometry.  Fortunately, the technologies needed for formation flying were picked up by other programs (see Figure~\ref{gallery}), either integrated into ``smallSat'' programs or as part of technology development for the LISA Gravitational Wave Interferometer Mission, which was recently given a huge boost by the detection of gravity waves by LIGO \citep{ligo2016}.   In the next section, we breakdown the subsystems needed for a formation flying space interferometer  and introduce some recent relevant smallSat and larger missions.

\newpage
\noindent {\bf Technology Drivers:}

\begin{wrapfigure}[18]{r}{0.6\textwidth}
\vspace{-.4in}
  \includegraphics[width=.6\textwidth]{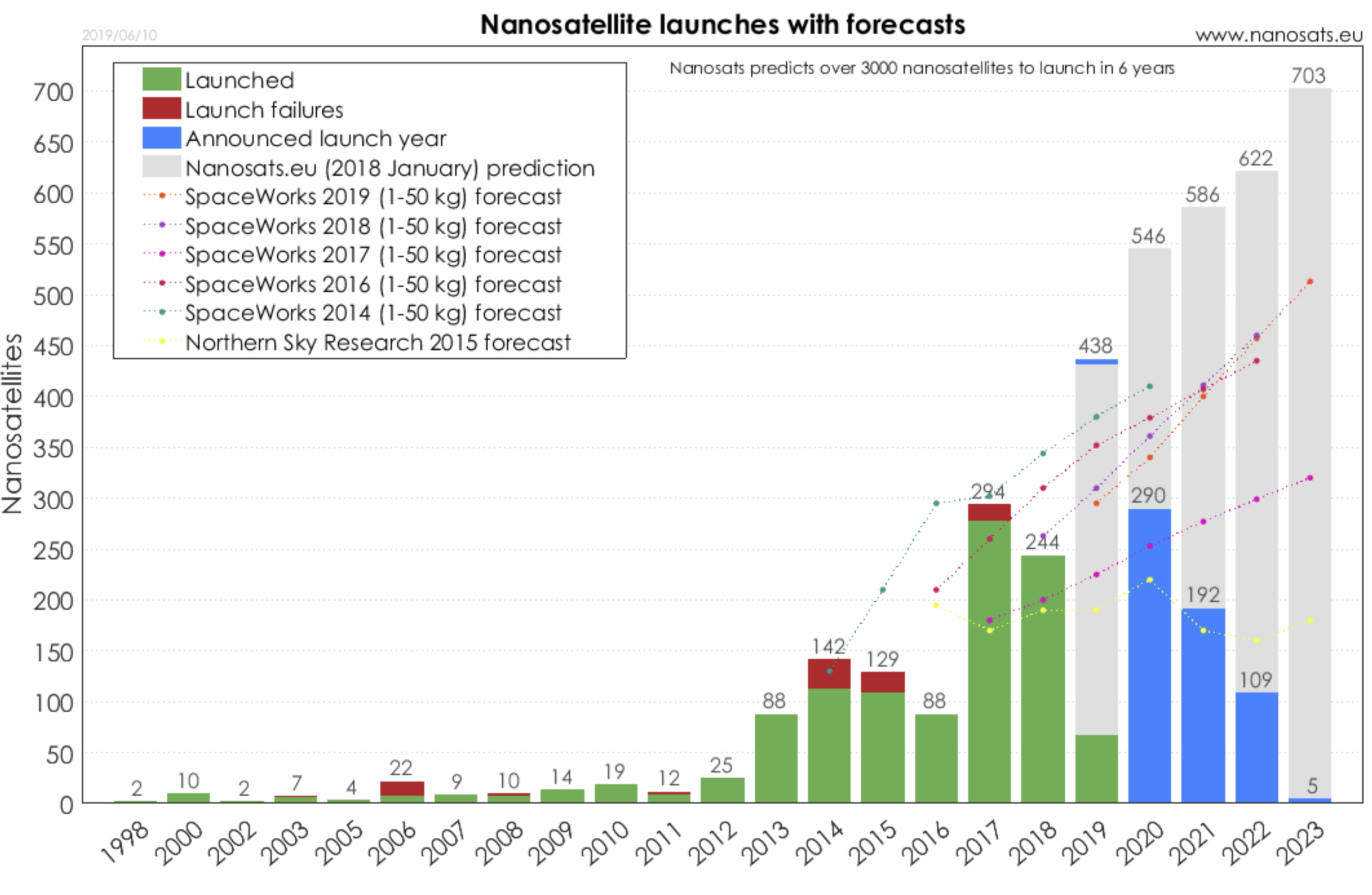}
\vspace{-.2in}
\caption{\footnotesize Around 1000 smallSats have flown so far with nearly 3000  in planning stages.  There is truly an engineering revolution happening surrounding inexpensive small satellites making formation flying space interferometry feasible and exciting to pursue over the coming decade. This graph was generated at http://www.nanosats.edu . }
\label{nanosat}       %
\end{wrapfigure}

The number of small satellite projects worldwide is skyrocketing.  Figure~\ref{nanosat} shows the incredible growth in the field with over 4000 missions launched or in planning stages. While most projects have nothing to do with astronomy directly,  astronomers have played major roles in some efforts, developing techniques for precision pointing and testing hardware such as cameras, detectors, fiber injection, and deformable mirrors.  In this section, we explore how formation flying space interferometry can leverage progress across a spectrum of technologies and be realistically pursued in this decade.

 It is beyond the scope of this white paper to go into detail about the Technical Readiness Level (TRL) of all required subsystems needed for a successful formation flying interferometer mission. However, we will discuss some of the key technologies and give a preliminary status report. We also present a reference list of relevant past, present and future projects (see Table~\ref{small_missions}).  Readers should please contact the lead author at monnier@umich.edu with corrections and additions, as we intend this document to stay updated and  available online following the ASTRO2020 process.

Astronomers might be curious why there has been so much recent technological development relevant to formation flying, outside the NASA/ESA interferometer context.  Telecommunication and remote sensing communities have interests in the same technologies as well as specific interests in operation of ``formations", ``swarms", and ``constellations'' of satellites, be they nanoSats, CubeSats, microSats, or larger spacecraft.  For example, absolute meter  station keeping is desired for orbit maintenance of even loose constellations, involving space GPS or ground referencing.  Basic maneuverability and propulsion systems are useful for changing orbits, re-pointing, and avoiding obstacles for a wide class of missions beyond strict formation flying.  LIDAR and radio ranging are technologies needed for mapping Earth's gravitational field or for future remote sensing missions, e.g., asteroids mapping, as well as for docking and rendezvous maneuvers.  

Flying spacecraft  within a formation $<$1~km across can in principle allow high-speed communication through 5G wireless networks which can enable radically different instrument architectures.  A ``fractionated spacecraft'' allows the different subsystems ordinarily contained on a single large, expensive spacecraft bus to be split into separate parts, allowing greater redundancy, cheaper replacements and sometimes fewer moving parts than traditional monolithic design.  For example, a sunshield used for cooling could be positioned separately, a calibration target could be flown  without needing awkward extending booms or other moving parts, and a high-gain antenna or high-bandwidth laser communication unit could be independent, relaying signals collected locally through the high speed local area network to ground stations. If a critical component fails, one can replace just the failed small satellite not an entire complex spacecraft.

Groups working with CubeSats have been pushing towards standardization of interfaces, underlying software architectures, miniaturization of key components such as propulsion valves, atomic clocks, LIDAR systems, etc.  {\bf A review of the field gives great optimism that most of the essential components for formation flying have been developed and tested in space already, although the need for consolidation and greater reliability remains. }

A realistic roadmap for formation flying space interferometry must include technology readiness in the following areas:

\begin{enumerate}
    \item Launch vehicles and available orbits. SmallSats are carried into space by a wide variety of launch vehicles and generally travel on low earth orbits (LEO) between 400-550km with orbital decay lifetimes between 1-10 years.  Wider orbits with longer lifetimes and higher orbital stability for the formation will ultimately be needed for space interferometry and additional study is needed to understand how to operate initial space interferometers from LEO.  There may be new launch opportunities for lifting smallSats to near-GEO or lunar orbit in the future that could be transformative for a formation flying mission.
    \item Post-release spacecraft stabilization (detumbling) and pointing acquisition.  Small satellites can be tumbling wildly following deployment.  Some mission failures still occur when trying to properly slow rotations, acquire a pointing reference and establish first ground communication.  This area is rapidly maturing and we can expect that reliable, off-the-shelf solutions should be available for any future formation flying mission.
    \item Communication with the ground.  Science programs often generate large volumes of data which need high bandwidth ground communication. Radio downlink during limited time periods per day is the general strategy but laser communication is possible with speeds up to $\sim$10Gbits/sec.  New planned satellite constellations (e.g., Starlink) providing high speed internet anywhere on (or just above) the Earth may be a game changer.
    \item Communication between spacecrafts in-formation.  Off-the-shelf Gigabit-speed local area networks based on 5G wireless technology are rolling out.  The km-level range is a good fit for formation flying interferometers.
    \item Power for smallSats is limited ($\sim$10W) due to the their small solar panels. More power will be needed for certain critical subsystems, e.g. laser communication and ranging.  Deployable panels with greater collecting areas are being developed commercially ($>40$W) and may mitigate these issues in the medium-term.
    \item Pointing, precision attitude control.  Maintaining stable and smooth pointing is critical for nearly all astronomical applications.  Recent missions have demonstrated arcsecond level pointing stability which should be (barely) adequate for formation flying of small apertures -- especially in conjunction with field steering mirror feedback. 
    \item Monitoring absolute and relative distance between spacecraft.  An overlapping hierarchy of technologies exist which allow us to know the 3D absolute and relative positions of a spacecraft formation.  Off-the-shelf Space-GPS can provide sub-meter absolute positions with cm-level relative positions, adequate for monitoring orbit and for collision avoidance.  For formation flying interferometry, we need to track separations within a  wavelength of light, well beyond mm ranging precision of state-of-the-art radio frequency ranging methods.  Laser ranging  has demonstrated spacecraft separations with precision down to the $\sim$10nm level and will be needed for formation flying interferometry.  
    \item Control and maneuverability, propulsion, orbit maintenance,  formation rigidity.  Propulsion is needed to change the baseline lengths between spacecrafts in an interferometer and also to maintain a fixed distance.  A serious issue for formation flying spacing interferometry involves how to maintain optical path differences to allow phase-stable, long coherent exposures of interference fringes.  
    Recent missions have maintained a formation with  $\pm$0.5~m separations using active thruster feedback and drift speeds of $\sim$100$\mu$m/sec have been achieved.  This performance is not adequate for formation flying interferometry and progress is needed in this area, including investigating chemical/monopropellant and electrical propulsion methods in addition to optimizing conventional cold gas thrusters.  We can expect  ongoing technical developments for LISA  and others missions to advance this general area.
    \item Re-positioning, re-fueling, rendezvous and docking.  In addition to using propellant to maintain the array  formation during observations, each new astronomical target will require globally re-orienting the array which is not possible with reaction wheels alone. The demands on propulsion will be high and limit both the achievable sensitivity and the lifetime of the mission.   Alternative propulsion methods such as using solar  pressure or atmospheric drag need to be explored for future formation flying.  In the longer term the ``propellant shortage problem'' can be solved through regular robotic refueling, thus requiring rendezvous and docking development generally not considered for smallSats but which can have applications for assembling space structures beyond formation flying interferometers.  Indeed, industry is pursuing on-orbit servicing capabilities both at GEO and LEO and we can reasonably expect remote re-fueling to be feasible on the decade time scale.
    \item Algorithms for situational awareness, autonomy, safety. Spacecraft distances within 100m are considered very risky and generally avoided.  If one loses control of a spacecraft in formation, there is a serious risk of collision and the generation of dangerous space debris.  Development of robust autonomous control algorithms and ``passively safe'' formations are essential for future formation flying interferometers.  On the software side there are connections with the active area of drone formation flying, situational awareness legacy from driverless cars, and past formation flying space missions.
    \item Reliability. While it is widely acknowledged that component reliability has improved significantly in recent years, formation flying space interferometry will place huge new demands as the required number of working subsystems stack up.
       \item General miniaturization and power reduction. University groups and companies are working to miniaturize a wide range of high-tech components that can be useful for modern instruments, for example lasers, atomic clocks, GPS hardware.  3D printing in plastic and metal has been a critical enabling technology for smallSats that allow optimal use of volume and mass and is accessible to small groups with limited budgets.   These advances are critical and affect the panoply of technologies listed above.   A future system-level review for a formation flying mission will need to identify key crucial technologies requiring further optimization. 
    \item Astronomy-specific challenges. So far we have hardly mentioned the unique astronomy requirements for formation flying space interferometry.  University groups can have a huge impact here, while counting on the legion (thousands!) of other smallSat groups to continue perfecting general-purpose capabilities. We need to identify solutions for telescopes, automated high-efficiency fiber coupling,  fine optical path length control to compensate for spacecraft drifts, beam steering and control, suitable ``space-friendly'' beam combiner designs, low power and space-qualified detectors,  compact and low power  cryogenics  for eventual infrared work,  sunshields for thermal management as well as control of scattered light, and evolve our thermal and vibration modelling capabilities.   We expect participation at smallSat sessions at the biennial SPIE meetings to swell over the coming decade.

\end{enumerate}

While smallSats can not accomplish all the ambitious science goals astronomers have, they offer a flexible and inexpensive prototype platform as a bridge to the \$1B missions we dream of. 
We will end this section with Table~\ref{small_missions} -- a partial list of mostly smallSat and some larger missions relevant to formation flying and space interferometry.  While out of scope to present full discussion of each project, we do include noteworthy details about each mission in the table and a paper reference when available.

\begin{table*}[!b]
\vspace{-.3in}
\footnotesize
\begin{center}
\caption{Missions developing enabling technology for formation flying space interferometry}
\begin{tabular}{|l|l|l|}
\hline
Name & Status & Primary Goal/s \\
\hline
Starlight (NASA-JPL) & cancelled 2002 & Formation flying space interferometer \\
&& \citep{blackwood2003} \\ 
TechSat-21 (AFRL) & cancelled 2003 &  Test formation flight technology \\
PRISMA (Sweden) & launched 2010 & Autonomous formation flying, 800m-5km,$\pm0.1m$ \\ 
&& \citep{persson2010} \\
TanDEM-X (Germany) & launch 2010 & Formation flying \citep{jaggi2012} \\
F6 (DARPA) & cancelled 2013 & Fractionated free-flying spacecraft (\$$>200$M spent) \\
CanX-4/5 (Canada) & launched 2014 & GPS  formation flying 50m-2300m separations  \\
&& Achieved positioning  $\pm$0.5m w/cold gas  \citep{kahr2018} \\
MMS (NASA) & launched 2015 & GPS-assisted formation flying; 4.5km apart \\
MinXSS (Colorado) & launched 2016 & Precision pointing \citep{mason2017} \\
OCSD-A (AeroSpace) & launched 2015 & Laser communication  \& \\
\& OCSD-B/C &launched  2017& Proximity sensing/maneuvering \citep{welle2018} \\
FLOCK 2p (Planet Lab) & launched 2017 & Constellation phasing with drag \citep{foster2018} \\
RANGE (Georgia Tech) & launched 2018 &  Laser ranging, controlling formation actively \\
PICSAT (Obs. Paris) & launched 2018 & Starlight injection into single mode-fiber; APD detector \\
& &\citep{nowak2018} \\
ASTERIA (MIT-JPL) & launched 2018 & Sub-arcsec pointing stability; $\Delta T\pm0.01$K \citep{pong2010}\\
GRACE-FO (NASA) & launched 2018 & Laser ranging interferometry over 230km  \\
&& with $<$10nm precision \citep{abich2019} \\
DeMi (MIT) & to launch 2019 & MEMS mirror testing \citep{allan2018} \\
TARGIT (Georgia Tech) & to launch 2020 & LIDAR, ranging measurement of two spacecraft \\
PROBA-3 (ESA) & to launch 2020 & Precision formation flying 150m apart \\
  && Goal $\pm$5cm transverse positioning \citep{focardi2015} \\ 
FIRST-S (Obs. Paris) & TBD $>$2022 & Interferometer on single spacecraft \citep{lapeyrere2018} \\
VTXO (NASA-GSFC) & TBD $>$2023 & Formation flying \citep{rankin2018} \\
mDOT (Stanford) & TBD $>$2025 & Formation flying \citep{koenig2015} \\
SunRise (NASA-JPL) & TBD $>$2025 & GPS formation flying with 6 spacecraft; \\
&& radio interferometry near GEO \citep{lazio2018} \\
LISA (ESA/NASA) & TBD $>$2034 & Formation flying 2.5 million km apart \\
& & \citep{danzmann2003}\\
\hline
\end{tabular}
\label{small_missions}
\end{center}
\vspace{-.2in}
\end{table*}

\newpage

\vspace{0.1in}
\noindent {\bf Organization, Partnerships, and Current Status:}

In addition to solving remaining technology issues, there are important management issues to consider.  Hardware and software expertise for smallSats is distributed amongst a diverse group of small university groups, government labs, industrial partners both large and small, and spread over the world.  Attempting formation flying space interferometry will require management of a loose-knit group of collaborators providing separate subsystems needing tight integration.  This is further complicated by intellectual property issues and government rules restricting space technology information sharing.  Formation flying space interferometry has a unique feature that helps organization since independent groups, even from different countries, can each be responsible for a separate spacecraft in the formation thus quarantining any IP and export licensing concerns to within each group.  As ground-based interferometry is already very international, we anticipate fruitful international collaborations formed to tackle the exciting challenge of formation flying space interferometry using small satellites.
We are aware of active interest to develop smallSat interferometry around the world, e.g., in the USA (Monnier), France (Lacour), and Australia (Ireland). 

 The  convergence of multiple developments -- exciting exoplanet science opportunities, the burgeoning smallSat revolution, LISA interferometer excitement, spiraling costs of NASA flagships -- has fueled renewed interest in space interferometry in the astronomical community. At the 2018 SPIE Meeting in Austin, a splinter group organized an impromptu working lunch to begin planning a new future for space interferometry.  This small group of $\sim$10 persons represented USA, European and Australian Universities along with NASA Centers, STScI, and the US aerospace industry.  In December 2018, the USRA organized a workshop with $\sim$40 participants to reboot long-stalled efforts, revisit the science cases for space interferometry, and brainstorm new ways of going forward. 

Beyond astronomy, the broader formation flying smallSat community is growing rapidly and with specialized workshops  devoted just to formation flying engineering.  We will start to see interferometrists at these meeting and begin converging on a baseline design for a first-generation formation flying space interferometer.

\vspace{.1in}
\noindent {\bf Schedule:}

First fringes using a formation flying space interferometer should be demonstrated by 2030.

\vspace{.1in}
\noindent {\bf Cost Estimates:}

For the topic of formation flying space interferometry, we are only proposing small satellite development over this decade, thus locating this effort firmly as a ``Small $<$\$500M'' space project.

{\em Thousands} of smallSat missions will occur over the next few years on a full range of topics, each developing important technology and maturing the smallSat ecosystem.  We anticipate at least 3 small cubeSat missions ($<$\$5M) will narrowly focus on crucial gap technology needed for formation flying during the first half of next decade, followed by a mid-scale (\$20-50M) mission to actually interfere light collected by separate spacecraft to be launched by 2030.

Funding for smallSats can come from a variety of places, including NASA,  NSF, DOD,  AFRL, and industry. Some known funding programs are listed in Table~\ref{funding}.  {\bf NASA and NSF should increase funding for smallSat astronomy, an investment that could lead to transformative capabilities for decades to come.}

\begin{table*}[!t]
\vspace{-.3in}
\footnotesize
\begin{center}
\caption{Relevant Funding Programs for Formation Flying Space Interferometry
(subject to change)}
\begin{tabular}{|l|l|l|l|}
\hline
Program Name & Purpose & Project Budget & Cadence (if known) \\
\hline
NASA ROSES/APRA & suborbital, CubeSats & \$5M, up to \$10M & yearly, March \\
NASA Mission of Opportunity (MOO) & SmallSats & \$35M, up to \$70M & every $\sim$2 years\\
NASA SMEX & small missions & \$180M w/ launch & every $\sim$4 years (next 2019) \\
NASA Explorers & small/medium missions & \$350M w/ launch & every $\sim$4 years (next 2021)\\
NASA Probes & large missions & \$1B & uncertain, $>$2022 \\
NASA SIMPLEx & small/planetary & \$20-\$50M &  \\
NASA ACT & components & \$1M-\$5M & \\
\hline
NSF Ideas Lab: Cross-cutting Initiative & CubeSat & \$4M & \\
 in CubeSat Innovations & & &\\
\hline
Air Force Research Lab & smallSats & 2 projects per year & next 2021 \\
University NanoSat Program & & & \\
\hline
\end{tabular}
\label{funding}
\end{center}
\vspace{-.2in}
\end{table*}

\clearpage
\pagebreak
\newpage
{\small
\bibliography{ASTRO2020_FFI}
\bibliographystyle{apalike}
}

\end{document}